\documentclass[modern,tighten]{aastex631}
\usepackage{amsmath}

\newcommand{\eb}{\begin{equation}}
\newcommand{\ee}{\end{equation}}

\newcommand{\uas}{\mbox{$\mu$as}}
\newcommand{\kms}{\mbox{km~s$^{-1}$}}

\usepackage{subfigure,graphicx}
\usepackage{url}
\usepackage{color}
\definecolor{darkgray}{gray}{0.4}
\definecolor{patinared}{rgb}{.72,.10,0}
\definecolor{patinablue}{rgb}{0,.20,.65} 
\definecolor{orange}{rgb}{1,0.5,0}
\definecolor{rkka}{RGB}{219,66,32}

\hypersetup{
  colorlinks=True,      
  urlcolor=darkgray,    
  linkcolor=patinared,  
  citecolor=patinablue, 
}
\makeatletter
\renewcommand{\frontmatter@title@above}{version \today \vskip 10mm}
\makeatother

\shorttitle{Extrapolation of Gaia parallax offset and its possible origin}
\shortauthors{Makarov \& Berghea}
\graphicspath{{./}{figures/}}

\begin{document}

\title{Gaia parallax bias via spherical harmonics: A Python tool and discussion of possible causes}

\email{valeri.v.makarov.civ@us.navy.mil}

\author[0000-0003-2336-7887]{Valeri V.\ Makarov}
\affiliation{U.S. Naval Observatory, 3450 Massachusetts Ave NW, Washington, DC 20392-5420, USA}
\author{Ciprian T.\ Berghea}
\affiliation{U.S. Naval Observatory, 3450 Massachusetts Ave NW, Washington, DC 20392-5420, USA}

\begin{abstract}
Parallaxes in Gaia DR3 are known to suffer from a complex set of sky-correlated and magnitude-dependent offsets or biases at the level of a few tens of $\mu$as. Estimated from a sample of one million distant quasars and AGNs from the CRF catalog, the average offset is negative, but the actual distribution of this important parameter shows significant variations on the sky. We propose a practical method to evaluate the parallax correction as a function of sky position and, optionally, of $G$ magnitude using a spherical harmonic series, and supply a tested Python tool {\tt varpi3.py} available on Zenodo\footnote{ https://zenodo.org/records/21708614}. We find that only the constant $Y_{00}$ term is significantly dependent on magnitude, while the other 80 harmonic terms are either close to zero or flat with magnitude. The directions of the smallest and largest parallax offsets are $(l,b)\simeq(220\degr,+43\degr)$ and $(l,b)\simeq(45\degr,-45\degr)$, which are close to the orientation of the quasar density dipole reported in recent publications. Motivated by this curious coincidence, we review possible physical effects resulting in a negative bias of measured parallaxes, including an anisotropic universe with a positive curvature and an orbital aberration component. The proposed method of parallax correction is tested using independent asteroseismology data for four different areas on the sphere. Finally, we show that the parallax zero-point propagates into the CRF proper-motion field through the parallax--proper-motion covariance, biasing the vector spherical harmonic determination of the secular-aberration glide, and hence the Galactocentric acceleration, at the microarcsecond-per-year level. 
\end{abstract}

\keywords{Astrometry(80), Gaia(2360), VLBI(1769), Astronomical techniques(1684), Proper motions(1295)}

\section{Introduction}
The problem of biases in the observed Gaia parallaxes has attracted considerable attention in the astronomical literature. The presence of such correlated errors in Early Data Release 3 \citep{2021A&A...649A...1G} and the current Data Release 3 was revealed by \citet{2021A&A...649A...2L} from an all-sky analysis of the CRF catalog comprising distant extragalactic AGNs and quasars. In fact, the necessity to evaluate the systematic errors is the main justification for obtaining the full 5- or 6-parameter astrometric solutions for CRF sources despite the strong prior of zero parallax for them. The average Gaia parallax of infinitely distant objects was found to be roughly $-20$ \uas. A systematic error of this kind arises from certain periodic variations of the basic angle \citep{2017A&A...603A..45B}, which are difficult to calibrate by the available technical means. A more detailed analysis of the parallax bias revealed a complex pattern of possible dependencies, which extends beyond the basic angle variation model \citep{2021A&A...649A...4L}. The averaged parallax error depends on the position of the source, its broadband magnitude $G$, and possibly color. \citet{2021A&A...649A...4L} offered two schemes ($Z_5$ and $Z_6$) to compute a posteriori corrections to the observed parallaxes from input coordinates, effective wavelength, and $G$ magnitude. In a series of subsequent publications, some limitations and shortcomings were revealed of the proposed correction techniques.

\citet{2021A&A...654A..20G}, using a combination of CRF and wide binary star samples, found that the color dependence of the parallax offset is weak, if any, and that proposed corrections based on sky position and magnitude for the brighter part of the test sample make the parallaxes consistent with independent data from Hubble Space Telescope (HST) measurements and classical cepheid models. \citet{2025AJ....169..211D}, using a combination of HST and VLBI data, detected a large negative offset of Gaia parallaxes at the brightest stellar magnitudes, which are not present in the CRF sample. A deeper negative offset with significant spatial variations was reported for the earlier DR2 on a relatively sparse sample of red-clump giants \citep{2020MNRAS.493.4367C}. Small samples of cepheids, HST targets, or radio stars in \citep{2022AstL...48..790B} are insufficient to probe the structured distribution of parallax offset with sufficient precision, however. Much larger stellar samples counting tens of thousands of sources reveal a complex pattern of the offset with deviant areas of the sky \citep{2021ApJ...910L...5H}. 

From the technical point of view, the proposed algorithms of correction based on the healpix partitioning of the sphere suffer from the artificial discretization noise. The main component of random error in this case follows a Poisson distribution defined by the number of data points in a given pixel that has a triangular shape. Since the offset determinations in adjacent partitions are independent, two nearby sources can obtain drastically different estimates if they happen to land in different partitions. There is an uneasy trade-off between smaller healpix areas to reveal the finer structure of the distribution and larger pixels to achieve some degree of consistency. Furthermore, a different scheme of geometric partitioning can provide a significantly different result for the same source. We advocate for a correction algorithm that controllably achieves some smoothing of the offset pattern, thus interpolating the fitted function between the available nodes and reducing the emergent Poisson noise.

The primary goal of this study is to provide a practical Python-based tool {\tt varpi3.py} (see Appendix for details of its operation) to compute the parallax correction as a function of sky coordinates over the entire celestial sphere and, optionally, of source magnitude. The offset-fitting method is based on the scalar spherical harmonic (SSH) decomposition of the Gaia-measured parallaxes of more than 1 million CRF objects, which provides the largest data sample and the most precise fit. The downside of this method is that it effectively covers only the range of optical quasars, i.e., between $G$ magnitudes 18 and 21, but a dedicated verification study confirms that the corrections work for systematically brighter stars with asteroseismological distances quite well. Inspired by a resemblance of the reconstructed map of parallax offset to the recently discovered anisotropy in the distribution of quasars on the sky, we also briefly review possible physical and cosmological origins of the persistent parallax bias.

\section{Selection of the data sample}
The main source of input data for this study is the collection of 1.6 million Gaia DR3 Celestial Reference Frame (CRF) sources with spectroscopic or Machine-Learning synthetic redshifts as described in \citet{2025NatAs...9.1396M}. This initial choice implies that the large majority of the sources are distant AGNs and quasars that are sufficiently bright in the optical and have infrared photometry from WISE and unWISE. We determined that additional quality filters were to be applied to the initial sample to drastically reduce the remaining contamination by stars. Even at a low rate of $<1\%$, stellar contaminants from distant parts of the Galaxy or nearby galaxies (e.g., dwarf satellites) can bias the parallax estimation toward positive values. A few additional selection cuts were applied to the 1.6 million CRF sources. The most radical selection removes all Gaia solutions with 6 fitting parameters (value 95 in {\tt astrometric\_params\_solved}). These special cases involve an additional ``pseudo-colour" parameter when the internal spectrophotometric determination could not be made, typically for marginally faint or otherwise perturbed images. The astrometric calibration procedures were more complex and less reliable for the 6-parameter solutions. As a result, these objects have a distinct set of systematic errors. We will focus on the more reliable and accurate 5-parameter solutions, taking into account that future Gaia data releases will not include any 6-parameter solutions.

Additionally, we have removed all solutions with {\tt ruwe}$<1.198$, which is the 0.97-quantile of the starting sample, and {\tt ipd\_gof\_harmonic\_amplitude}$<0.2$. The latter quality parameter was designed to capture unresolved binary images by analyzing the shape of the image at different scan directions. Because of an implementation error in DR3, it is instead strongly correlated with photometric variability of the given source. We still consider it to be useful to remove potential perturbers in the global fit of parallax. These cuts reduced the starting sample to 1.157 million sources. 

\begin{figure}
\includegraphics[width=1\textwidth]{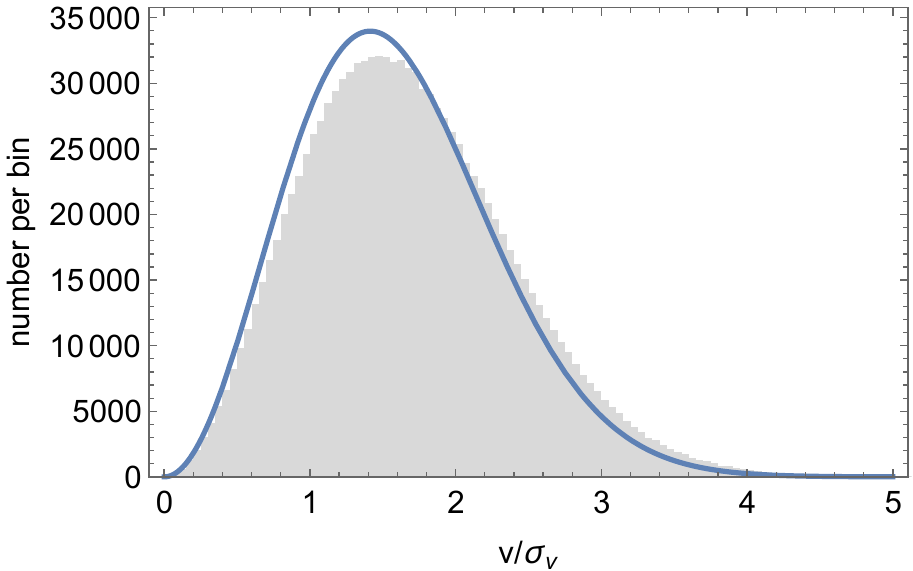}
\caption{Histogram of standardized parallax-proper motion vectors $\sqrt{w}\equiv v/\sigma_v$ for 1.157 million Gaia CRF sources in the working sample after preliminary quality cuts. The blue curve shows the expected $\chi[3]$ distribution.}
\label{chi.fig}
\end{figure}

To quantify the overall astrometric quality of the working sample, we computed a combined statistics on the observed 3-vector $\boldsymbol{v}=[\varpi, \mu_{\alpha*},\mu_\delta]^T$, which includes the parallax and proper motion components for each source. All these values should be equal to zero within the measurement precision for extragalactic objects. Therefore, the quadratic form 
\eb 
w\equiv \boldsymbol{v}^T\,\boldsymbol{C}_3^{-1}\;\boldsymbol{v}
\ee 
is expected to be $\chi^2[3]$-distributed assuming Gaussian probability densities for the three involved measurements. $\boldsymbol{C}_3$ is the formal $3\times 3$ covariance matrix constructed from the data given in the catalog for each source. The square root of $w$ can be considered as a signal-to-noise ratio (SNR)  for the combined parallax and proper motion measurement. This statistic is more powerful than the separate $\chi^2$-based indicators for parallax and proper motions, because it takes into account the known covariances between these values. Fig. \ref{chi.fig} shows the histogram of $\sqrt{w}$ values (labeled as $v/\sigma_v$) after the initial quality cuts. The expected $\chi[3]$ distribution is shown with a blue line. We can see that the scatter of $\boldsymbol{v}$-vectors is still larger than the theoretical prediction, which is reflected in a small upward shift of the mode and excess rate of values in the tail. These effects can be caused by the presence of stellar contaminants with finite parallaxes and proper motions, although slightly underestimated formal errors or occasional image perturbations cannot be precluded. To be on the safe side, we additionally applied a filter $v/\sigma_v<2.5$, ending up with a sample of 1.004 million sources.

\section{Data analysis}
\label{meth.sec}

The initial step of data analysis included sorting the filtered sample of CRF sources by $G$ magnitude given in the catalog. The sorted sample was subsequently divided into 10 equal batches of 100,370 sources each. This partitioning provides sufficient statistical reliability of the estimated parallax offsets while sampling the entire range of magnitudes. 

The ICRS coordinates of the sources were transformed into galactic coordinates following the standard algorithm \citep{1997ESASP1200.....E}. For each magnitude batch, we performed a fit of the measured parallaxes $\varpi$ as a function of galactic $\{l,b\}$ using scalar spherical harmonic (SSH) functions. Specifically, the fit is
\eb 
\varpi(l,b) = \sum_{n,m} a_{nm} Y_{nm}(l,b),
\label{obs.eq}
\ee 
where $a_{nm}$ are the free fitting coefficients, $Y_{nm}$ are the standard SSH functions\footnote{We used {\tt SphericalHarmonicY[n,m,l,b+Pi/2]} functions in Wolfram Mathematica and {\tt sph\_harm(m,n,l,b+np.pi/2)} in scipy.special} of degree $n$ and order $m$. For the complex-valued SSH, both their real and imaginary parts should be used as real-valued basis functions. The sum is truncated at a chosen degree $N$. We experimented with different values of $N$ and determined that $N=8$ (81 fitting functions) provides the best balance between the angular resolution of spatial distribution and precision of the fit. We also found out that SSH-fitting in galactic coordinates provides more stable and reliable results than doing this in the original equatorial coordinates. The reason for this dependence is the wide zone of avoidance around the Galactic plane where quasars are practically absent due to interstellar extinction and crowding. The incomplete sky coverage generates internal correlations between the fitting functions and results in a loss of condition, which can be quantified using the SVD-based robustness parameter \citep{2021AJ....161..289M}. The complete list of SSH functions with the adopted normalization coefficients is given in Table \ref{sshlist.tab}.

\newpage
\startlongtable
\begin{deluxetable*}{L|L}
\tablecaption{SSH functions to degree 8.}
 \label{sshlist.tab}
\tablehead{
\colhead{\text{name}} & \colhead{\text{SSH function}}
}
\startdata
 \{\text{re},0,0\} & \frac{1}{2 \sqrt{\pi }} \\
 \{\text{re},1,0\} & -\frac{1}{2} \sqrt{\frac{3}{\pi }} \sin (b) \\
 \{\text{re},1,1\} & -\frac{1}{2} \sqrt{\frac{3}{2 \pi }} \cos (b) \cos (l) \\
 \{\text{im},1,1\} & -\frac{1}{2} \sqrt{\frac{3}{2 \pi }} \cos (b) \sin (l) \\
 \{\text{re},2,0\} & \frac{1}{4} \sqrt{\frac{5}{\pi }} \left(3 \sin ^2(b)-1\right) \\
 \{\text{re},2,1\} & \frac{1}{2} \sqrt{\frac{15}{2 \pi }} \sin (b) \cos (b) \cos (l) \\
 \{\text{im},2,1\} & \frac{1}{2} \sqrt{\frac{15}{2 \pi }} \sin (b) \cos (b) \sin (l) \\
 \{\text{re},2,2\} & \frac{1}{4} \sqrt{\frac{15}{2 \pi }} \cos ^2(b) \cos (2 l) \\
 \{\text{im},2,2\} & \frac{1}{4} \sqrt{\frac{15}{2 \pi }} \cos ^2(b) \sin (2 l) \\
 \{\text{re},3,0\} & -\frac{1}{4} \sqrt{\frac{7}{\pi }} \sin (b) \left(5 \sin ^2(b)-3\right) \\
 \{\text{re},3,1\} & -\frac{1}{8} \sqrt{\frac{21}{\pi }} \left(5 \sin ^2(b)-1\right) \cos (b) \cos (l) \\
 \{\text{im},3,1\} & -\frac{1}{8} \sqrt{\frac{21}{\pi }} \left(5 \sin ^2(b)-1\right) \cos (b) \sin (l) \\
 \{\text{re},3,2\} & -\frac{1}{4} \sqrt{\frac{105}{2 \pi }} \sin (b) \cos ^2(b) \cos (2 l) \\
 \{\text{im},3,2\} & -\frac{1}{4} \sqrt{\frac{105}{2 \pi }} \sin (b) \cos ^2(b) \sin (2 l) \\
 \{\text{re},3,3\} & -\frac{1}{8} \sqrt{\frac{35}{\pi }} \cos ^3(b) \cos (3 l) \\
 \{\text{im},3,3\} & -\frac{1}{8} \sqrt{\frac{35}{\pi }} \cos ^3(b) \sin (3 l) \\
 \{\text{re},4,0\} & \frac{3}{16 \sqrt{\pi }} \left(35 \sin ^4(b)-30 \sin ^2(b)+3\right) \\
 \{\text{re},4,1\} & \frac{3}{8} \sqrt{\frac{5}{\pi }} \sin (b) \left(7 \sin ^2(b)-3\right) \cos (b) \cos (l) \\
 \{\text{im},4,1\} & \frac{3}{8} \sqrt{\frac{5}{\pi }} \sin (b) \left(7 \sin ^2(b)-3\right) \cos (b) \sin (l) \\
 \{\text{re},4,2\} & \frac{3}{8} \sqrt{\frac{5}{2 \pi }} \left(7 \sin ^2(b)-1\right) \cos ^2(b) \cos (2 l) \\
 \{\text{im},4,2\} & \frac{3}{8} \sqrt{\frac{5}{2 \pi }} \left(7 \sin ^2(b)-1\right) \cos ^2(b) \sin (2 l) \\
 \{\text{re},4,3\} & \frac{3}{8} \sqrt{\frac{35}{\pi }} \sin (b) \cos ^3(b) \cos (3 l) \\
 \{\text{im},4,3\} & \frac{3}{8} \sqrt{\frac{35}{\pi }} \sin (b) \cos ^3(b) \sin (3 l) \\
 \{\text{re},4,4\} & \frac{3}{16} \sqrt{\frac{35}{2 \pi }} \cos ^4(b) \cos (4 l) \\
 \{\text{im},4,4\} & \frac{3}{16} \sqrt{\frac{35}{2 \pi }} \cos ^4(b) \sin (4 l) \\
 \{\text{re},5,0\} & -\frac{1}{16} \sqrt{\frac{11}{\pi }} \sin (b) \left(63 \sin ^4(b)-70 \sin ^2(b)+15\right) \\
 \{\text{re},5,1\} & -\frac{1}{16} \sqrt{\frac{165}{2 \pi }} \left(21 \sin ^4(b)-14 \sin ^2(b)+1\right) \cos (b) \cos (l) \\
 \{\text{im},5,1\} & -\frac{1}{16} \sqrt{\frac{165}{2 \pi }} \left(21 \sin ^4(b)-14 \sin ^2(b)+1\right) \cos (b) \sin (l) \\
 \{\text{re},5,2\} & -\frac{1}{8} \sqrt{\frac{1155}{2 \pi }} \sin (b) \left(3 \sin ^2(b)-1\right) \cos ^2(b) \cos (2 l) \\
 \{\text{im},5,2\} & -\frac{1}{8} \sqrt{\frac{1155}{2 \pi }} \sin (b) \left(3 \sin ^2(b)-1\right) \cos ^2(b) \sin (2 l) \\
 \{\text{re},5,3\} & -\frac{1}{32} \sqrt{\frac{385}{\pi }} (3 \sin (b)-1) (3 \sin (b)+1) \cos ^3(b) \cos (3 l) \\
 \{\text{im},5,3\} & -\frac{1}{32} \sqrt{\frac{385}{\pi }} (3 \sin (b)-1) (3 \sin (b)+1) \cos ^3(b) \sin (3 l) \\
 \{\text{re},5,4\} & -\frac{3}{16} \sqrt{\frac{385}{2 \pi }} \sin (b) \cos ^4(b) \cos (4 l) \\
 \{\text{im},5,4\} & -\frac{3}{16} \sqrt{\frac{385}{2 \pi }} \sin (b) \cos ^4(b) \sin (4 l) \\
 \{\text{re},5,5\} & -\frac{3}{32} \sqrt{\frac{77}{\pi }} \cos ^5(b) \cos (5 l) \\
 \{\text{im},5,5\} & -\frac{3}{32} \sqrt{\frac{77}{\pi }} \cos ^5(b) \sin (5 l) \\
 \{\text{re},6,0\} & \frac{1}{32} \sqrt{\frac{13}{\pi }} \left(231 \sin ^6(b)-315 \sin ^4(b)+105 \sin ^2(b)-5\right) \\
 \{\text{re},6,1\} & \frac{1}{16} \sqrt{\frac{273}{2 \pi }} \sin (b) \left(33 \sin ^4(b)-30 \sin ^2(b)+5\right) \cos (b) \cos (l)
   \\
 \{\text{im},6,1\} & \frac{1}{16} \sqrt{\frac{273}{2 \pi }} \sin (b) \left(33 \sin ^4(b)-30 \sin ^2(b)+5\right) \cos (b) \sin (l)
   \\
 \{\text{re},6,2\} & \frac{1}{64} \sqrt{\frac{1365}{\pi }} \left(33 \sin ^4(b)-18 \sin ^2(b)+1\right) \cos ^2(b) \cos (2 l) \\
 \{\text{im},6,2\} & \frac{1}{64} \sqrt{\frac{1365}{\pi }} \left(33 \sin ^4(b)-18 \sin ^2(b)+1\right) \cos ^2(b) \sin (2 l) \\
 \{\text{re},6,3\} & \frac{1}{32} \sqrt{\frac{1365}{\pi }} \sin (b) \left(11 \sin ^2(b)-3\right) \cos ^3(b) \cos (3 l) \\
 \{\text{im},6,3\} & \frac{1}{32} \sqrt{\frac{1365}{\pi }} \sin (b) \left(11 \sin ^2(b)-3\right) \cos ^3(b) \sin (3 l) \\
 \{\text{re},6,4\} & \frac{3}{32} \sqrt{\frac{91}{2 \pi }} \left(11 \sin ^2(b)-1\right) \cos ^4(b) \cos (4 l) \\
 \{\text{im},6,4\} & \frac{3}{32} \sqrt{\frac{91}{2 \pi }} \left(11 \sin ^2(b)-1\right) \cos ^4(b) \sin (4 l) \\
 \{\text{re},6,5\} & \frac{3}{32} \sqrt{\frac{1001}{\pi }} \sin (b) \cos ^5(b) \cos (5 l) \\
 \{\text{im},6,5\} & \frac{3}{32} \sqrt{\frac{1001}{\pi }} \sin (b) \cos ^5(b) \sin (5 l) \\
 \{\text{re},6,6\} & \frac{1}{64} \sqrt{\frac{3003}{\pi }} \cos ^6(b) \cos (6 l) \\
 \{\text{im},6,6\} & \frac{1}{64} \sqrt{\frac{3003}{\pi }} \cos ^6(b) \sin (6 l) \\
 \{\text{re},7,0\} & -\frac{1}{32} \sqrt{\frac{15}{\pi }} \sin (b) \left(429 \sin ^6(b)-693 \sin ^4(b)+315 \sin ^2(b)-35\right) \\
 \{\text{re},7,1\} & -\frac{1}{64} \sqrt{\frac{105}{2 \pi }} \left(429 \sin ^6(b)-495 \sin ^4(b)+135 \sin ^2(b)-5\right) \cos (b)
   \cos (l) \\
 \{\text{im},7,1\} & -\frac{1}{64} \sqrt{\frac{105}{2 \pi }} \left(429 \sin ^6(b)-495 \sin ^4(b)+135 \sin ^2(b)-5\right) \cos (b)
   \sin (l) \\
 \{\text{re},7,2\} & -\frac{3}{64} \sqrt{\frac{35}{\pi }} \sin (b) \left(143 \sin ^4(b)-110 \sin ^2(b)+15\right) \cos ^2(b) \cos (2
   l) \\
 \{\text{im},7,2\} & -\frac{3}{64} \sqrt{\frac{35}{\pi }} \sin (b) \left(143 \sin ^4(b)-110 \sin ^2(b)+15\right) \cos ^2(b) \sin (2
   l) \\
 \{\text{re},7,3\} & -\frac{3}{64} \sqrt{\frac{35}{2 \pi }} \left(143 \sin ^4(b)-66 \sin ^2(b)+3\right) \cos ^3(b) \cos (3 l) \\
 \{\text{im},7,3\} & -\frac{3}{64} \sqrt{\frac{35}{2 \pi }} \left(143 \sin ^4(b)-66 \sin ^2(b)+3\right) \cos ^3(b) \sin (3 l) \\
 \{\text{re},7,4\} & -\frac{3}{32} \sqrt{\frac{385}{2 \pi }} \sin (b) \left(13 \sin ^2(b)-3\right) \cos ^4(b) \cos (4 l) \\
 \{\text{im},7,4\} & -\frac{3}{32} \sqrt{\frac{385}{2 \pi }} \sin (b) \left(13 \sin ^2(b)-3\right) \cos ^4(b) \sin (4 l) \\
 \{\text{re},7,5\} & -\frac{3}{64} \sqrt{\frac{385}{2 \pi }} \left(13 \sin ^2(b)-1\right) \cos ^5(b) \cos (5 l) \\
 \{\text{im},7,5\} & -\frac{3}{64} \sqrt{\frac{385}{2 \pi }} \left(13 \sin ^2(b)-1\right) \cos ^5(b) \sin (5 l) \\
 \{\text{re},7,6\} & -\frac{3}{64} \sqrt{\frac{5005}{\pi }} \sin (b) \cos ^6(b) \cos (6 l) \\
 \{\text{im},7,6\} & -\frac{3}{64} \sqrt{\frac{5005}{\pi }} \sin (b) \cos ^6(b) \sin (6 l) \\
 \{\text{re},7,7\} & -\frac{3}{64} \sqrt{\frac{715}{2 \pi }} \cos ^7(b) \cos (7 l) \\
 \{\text{im},7,7\} & -\frac{3}{64} \sqrt{\frac{715}{2 \pi }} \cos ^7(b) \sin (7 l) \\
 \{\text{re},8,0\} & \frac{1}{256} \sqrt{\frac{17}{\pi }} \left(6435 \sin ^8(b)-12012 \sin ^6(b)+6930 \sin ^4(b)-1260 \sin
   ^2(b)+35\right) \\
 \{\text{re},8,1\} & \frac{3}{64} \sqrt{\frac{17}{2 \pi }} \sin (b) \left(715 \sin ^6(b)-1001 \sin ^4(b)+385 \sin ^2(b)-35\right)
   \cos (b) \cos (l) \\
 \{\text{im},8,1\} & \frac{3}{64} \sqrt{\frac{17}{2 \pi }} \sin (b) \left(715 \sin ^6(b)-1001 \sin ^4(b)+385 \sin ^2(b)-35\right)
   \cos (b) \sin (l) \\
 \{\text{re},8,2\} & \frac{3}{128} \sqrt{\frac{595}{\pi }} \left(143 \sin ^6(b)-143 \sin ^4(b)+33 \sin ^2(b)-1\right) \cos ^2(b)
   \cos (2 l) \\
 \{\text{im},8,2\} & \frac{3}{128} \sqrt{\frac{595}{\pi }} \left(143 \sin ^6(b)-143 \sin ^4(b)+33 \sin ^2(b)-1\right) \cos ^2(b)
   \sin (2 l) \\
 \{\text{re},8,3\} & \frac{1}{64} \sqrt{\frac{19635}{2 \pi }} \sin (b) \left(39 \sin ^4(b)-26 \sin ^2(b)+3\right) \cos ^3(b) \cos
   (3 l) \\
 \{\text{im},8,3\} & \frac{1}{64} \sqrt{\frac{19635}{2 \pi }} \sin (b) \left(39 \sin ^4(b)-26 \sin ^2(b)+3\right) \cos ^3(b) \sin
   (3 l) \\
 \{\text{re},8,4\} & \frac{3}{128} \sqrt{\frac{1309}{2 \pi }} \left(65 \sin ^4(b)-26 \sin ^2(b)+1\right) \cos ^4(b) \cos (4 l) \\
 \{\text{im},8,4\} & \frac{3}{128} \sqrt{\frac{1309}{2 \pi }} \left(65 \sin ^4(b)-26 \sin ^2(b)+1\right) \cos ^4(b) \sin (4 l) \\
 \{\text{re},8,5\} & \frac{3}{64} \sqrt{\frac{17017}{2 \pi }} \sin (b) \left(5 \sin ^2(b)-1\right) \cos ^5(b) \cos (5 l) \\
 \{\text{im},8,5\} & \frac{3}{64} \sqrt{\frac{17017}{2 \pi }} \sin (b) \left(5 \sin ^2(b)-1\right) \cos ^5(b) \sin (5 l) \\
 \{\text{re},8,6\} & \frac{1}{128} \sqrt{\frac{7293}{\pi }} \left(15 \sin ^2(b)-1\right) \cos ^6(b) \cos (6 l) \\
 \{\text{im},8,6\} & \frac{1}{128} \sqrt{\frac{7293}{\pi }} \left(15 \sin ^2(b)-1\right) \cos ^6(b) \sin (6 l) \\
 \{\text{re},8,7\} & \frac{3}{64} \sqrt{\frac{12155}{2 \pi }} \sin (b) \cos ^7(b) \cos (7 l) \\
 \{\text{im},8,7\} & \frac{3}{64} \sqrt{\frac{12155}{2 \pi }} \sin (b) \cos ^7(b) \sin (7 l) \\
 \{\text{re},8,8\} & \frac{3}{256} \sqrt{\frac{12155}{2 \pi }} \cos ^8(b) \cos (8 l) \\
 \{\text{im},8,8\} & \frac{3}{256} \sqrt{\frac{12155}{2 \pi }} \cos ^8(b) \sin (8 l) \\
\enddata
\end{deluxetable*}

We have produced both unweighted and weighted least-squares solutions of Eqs. \ref{obs.eq}, which yielded rather close, but not identical, estimates for the coefficients $a_{nm}$. The weighted option was chosen because it provided more realistic formal errors of the fitting coefficients as functions of magnitude. As usual in the assumption of a normally distributed observational error, the weight for source $i$ was set to $1/\sigma_{\varpi i}$. As a result of this least-squares adjustment, we determined a set of 81 SSH coefficients $a_{nm}$ and their formal errors std$[a_{nm}]$ for each of the 10 magnitude bins.

\section{Results and a correction model setup}
\label{res.sec}
\begin{figure}
\includegraphics[width=1\textwidth]{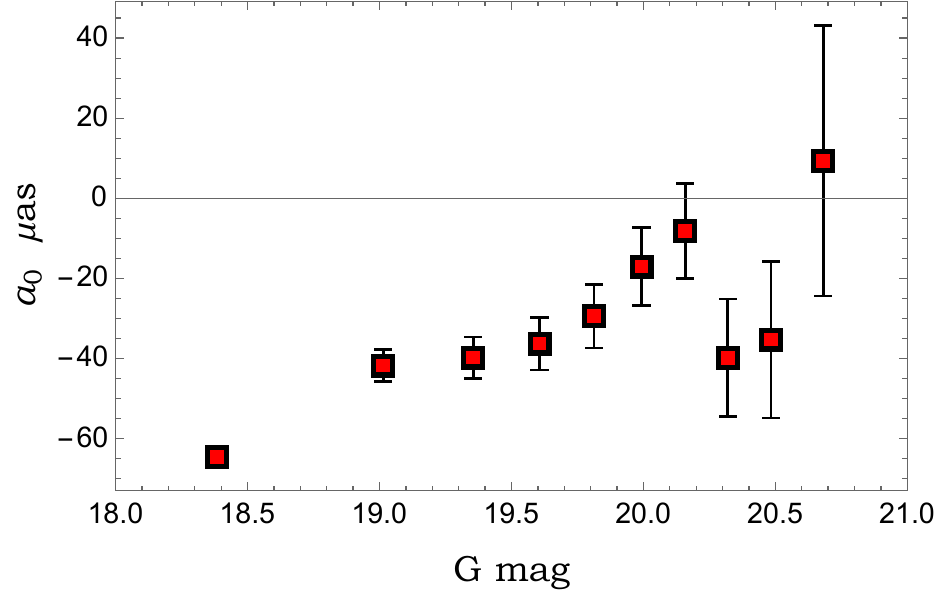}
\caption{Zeroth degree SSH coefficients $a_{00}$ (constant offsets) of Gaia DR3 parallax computed for 10 batches of CRF sources sorted by $G$ magnitudes. The median $G$ magnitudes are used as abscissae for the dots. The error bars represent the standard deviations extracted from the covariance matrices of the fits.}
\label{y00.fig}
\end{figure}

Fig. \ref{y00.fig} shows the fitted zero-degree coefficients $a_{00}$ with their standard errors for the 10 batches of $G$ magnitude. Each square indicates the estimated value for the median magnitudes within one batch. Since the uncertainties are obtained from the weighted covariance matrices in the least-squares fit, they are strongly magnitude-dependent. For the brightest batches down to $G=19.8$ mag, the detected harmonics are highly significant. We can see a nontrivial behavior of the general parallax bias with magnitude. The largest offset is found for the brightest sources at $a_{00}=-64$ \uas, which corresponds to a bias of $-18.2$ \uas\ (note that the fitted coefficients should be multiplied by the normalization coefficients given in Table~1 to convert to pure functional forms). Beyond the CRF cutoff at $G\simeq 18.4$ mag, \citet{2025AstL...51..516B} estimated an average bias of $-38$ \uas\ for brighter radio stars and masers. The sky-average offset seems to decrease for fainter sources, possibly nullifying at the end of the range around $G\simeq 20.7$.

\begin{figure}
\includegraphics[width=0.32\textwidth]{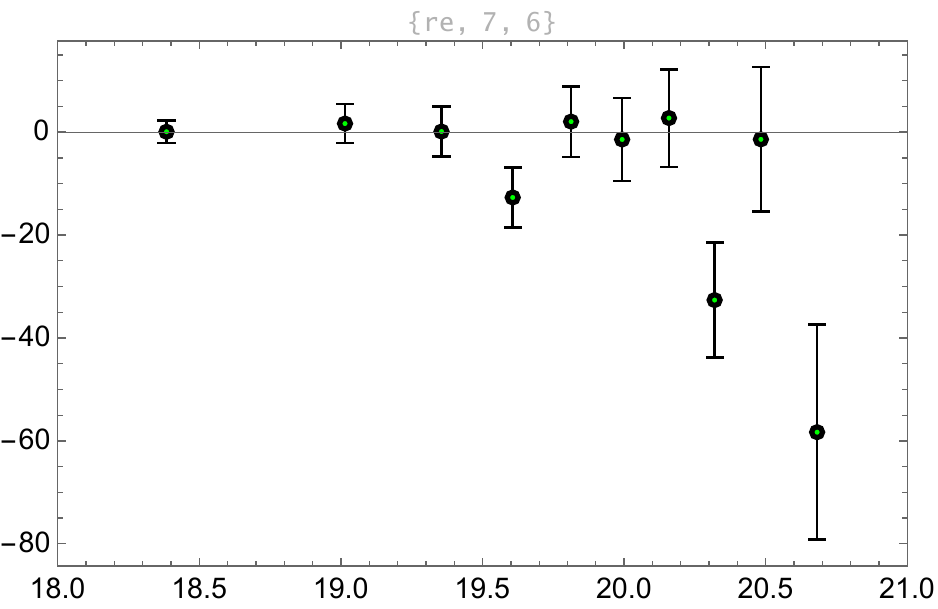}
\includegraphics[width=0.32\textwidth]{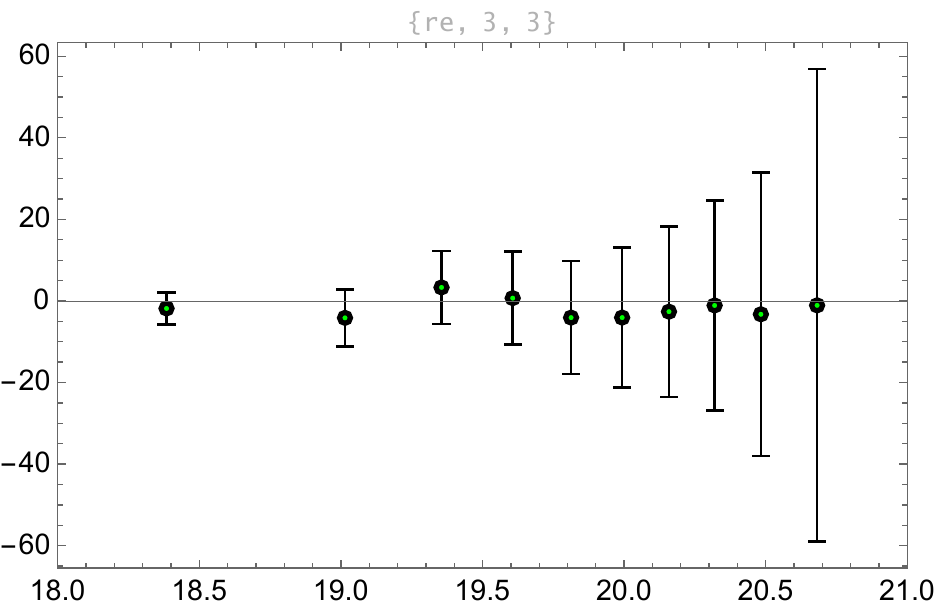}
\includegraphics[width=0.32\textwidth]{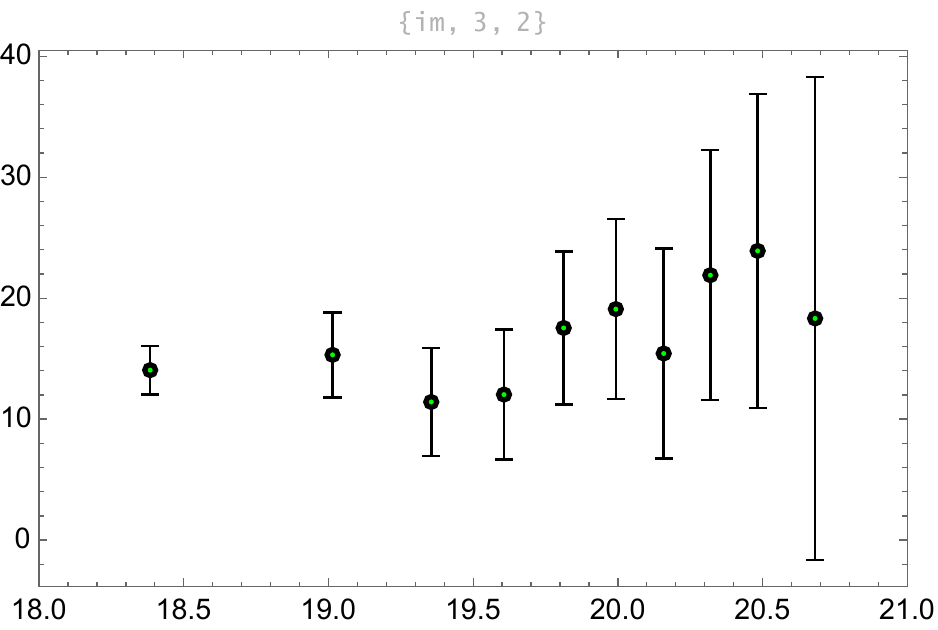}
\caption{Estimated SSH coefficients $a_{nm}$ in \uas\ for ten batches of $G$ magnitude. Only three representative examples are shown for SSH terms \{re,7,6\}, \{re,3,3\}, and \{im,3,2\}, as labeled on top of each plot. }
\label{3ex.fig}
\end{figure}

Similar plots were generated and examined for the other 80 spherical harmonics. Fig. \ref{3ex.fig} displays three typical cases of $a_{nm}(G)$ behavior. The fitted dependence $a_{76}(G)$ in the left plot is the most frequent outcome, where the estimates for the brighter objects are close to zero, but some variation seems to be present across the range of magnitudes, which is statistically insignificant, however. A few terms show remarkably flat and nearly zero behavior as in the middle plot representing $a_{33}(G)$. The right plot displays a rare case of a flat but certainly nonzero dependence. These terms are important for a realistic and accurate correction model, because the common (magnitude-independent) patterns of parallax bias cannot be neglected. 

To identify the terms where a statistically significant signal is present, we computed the weighted mean of each coefficient over the magnitude batches and their standard errors. The ratio of these values is a signal-to-noise estimate (SNR). Only 6 out of 81 terms emerge with SNR values exceeding 3.5, namely, \{re,0,0\} (SNR$=30.6$), \{re,1,1\} (6.4), \{im,1,1\} (7.0), \{im,2,1\} (4.9), \{im,3,2\} (10.2), and \{im,7,1\} (3.6). We note that the significant terms are mostly present in the low degrees, which reflects a red spectrum of parallax perturbations, i.e., a degree of smoothness in their distribution. Arguably, one can limit the decomposition to $N=7$, but numerical trials revealed that the resulting fit is not much different from the nominal $N=8$ case. The mean coefficients are given in the input table supporting the {\tt varpi3.py} tool.

We further quantified the significance of apparent variations of the fitting coefficients with $G$ for each SSH term using Pearson $\chi^2$-test on the error-normalized residuals after the weighted mean values had been subtracted. The emerging $p$-value in that test estimates is the probability, under the null hypothesis of normally distributed residuals, of obtaining a $\chi^2$ value at least as large as observed. Two terms out of 81 obtain $p$-values below 0.1, namely, \{re,0,0\} and \{re,7,6\}, while the expected number of random occurrences of such values is 0.81 due to the look-elsewhere effect, i.e., from testing 81 terms simultaneously. This result justifies the adopted strategy of parallax correction in the {\tt varpi3.py} tool. The magnitude dependence of SSH fitting coefficients is ignored for all terms except \{re,0,0\}, and the mean coefficients are used to compute the parallax correction for a given source. If a $G$-magnitude is given in the input for the tool, the \{re,0,0\} coefficient corresponding to the nearest nodal value of magnitude (see Fig. \ref{y00.fig}) is used. Effectively, the algorithm performs a zeroth-order interpolation/extrapolation of the constant term as an option chosen by the user. All the other SSH terms are assumed to be magnitude-independent.

\begin{figure}
\includegraphics[width=0.82\textwidth]{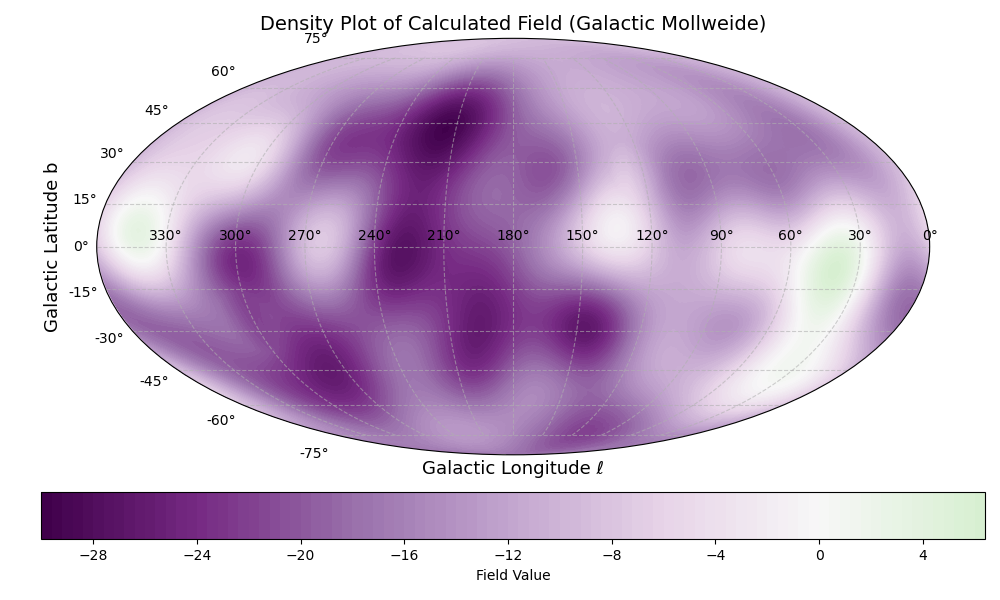}
\caption{Sky distribution of Gaia CRF3 parallax bias reconstructed with 81 spherical harmonic functions to degree 8. The color-coded offset values are in $\mu$as.}
\label{map.fig}
\end{figure}

Fig. \ref{map.fig} shows the distribution of parallax $\varpi$ reconstructed with the magnitude-averaged SSH coefficients (including the mean $\bar a_{00}$) for a uniform grid of $(l,b)$ points. The Galactic center in this plot is at the right margin of the projection. The color-coded parallax values reveal a structured global distribution with a range between $-30$ and $+5$ \uas, approximately. By design, this result is mostly defined by the brighter CRF sources, which have higher weight in the mean coefficients, and it should be valid for stars as well, which are often brighter than the average CRF quasar. We find a few prominent features representing relatively compact areas of negative and positive deviations from the zero-point bias. Some of these peaks and dips are aligned with the Galactic plane. These should be taken with caution, because the SSH fit is a mere extrapolation for the exclusion band. Indeed, the number density distribution of Gaia CRF3 sources (not shown for brevity) features a zone of exclusion at $|b|\lesssim 10\degr$. The gap causes a partial loss of condition in the global SSH fit, which mostly affects high-degree zonal harmonics. The peaks outside the exclusion zone are important, because they are well-conditioned and reliable. 

To identify the main structures more reliably, we reconstructed the $\varpi$ bias field using only 10 SSH terms with SNR$>2.5$. The resulting plot, which is a smoother version of Fig. \ref{map.fig}, reveals an asymmetric pattern with an area of the most negative parallax ($-27$ \uas) centered on $(l,b)\simeq(220\degr,+43\degr)$, and an area of the least negative bias ($-2$ \uas) around $(l,b)\simeq(45\degr,-45\degr)$. Note that the two directions are almost exactly opposite, and the negative peak has ecliptic coordinates $\lambda=142\degr$, $\beta=-1\degr$. Thus, the most prominent parallax deviations are seen in the ecliptic plane. The combined effect can be seen as a dipole perturbation with an axis in the ecliptic plane at an angle of $142\degr$ to the vernal equinox. The second lowest dip in average parallax outside the Galactic plane is found at approximately $(l,b)\simeq(280\degr,-42\degr)$.

Since the main features of the fitted parallax offset field form a structure resembling a dipole, and the explored objects are quasars at cosmological distances, it may be of interest to compare the emerging effect with the possible anisotropy of the global distribution of quasars. Using a large sample of quasars identified from all-sky infrared surveys, \citet{2021ApJ...908L..51S, 2022ApJ...937L..31S} found a pronounced asymmetry in the number counts in the direction of the CMB dipole exceeding the expected Doppler-boosting effect by a factor of 2. The CMB brightness dipole in the direction $(l,b)=(264\degr,48\degr)$ is caused by the peculiar velocity of the observer of 370 km s$^{-1}$ in this direction in the preferred inertial reference frame. Subsequent analyses refined and generalized this number density method, including treating the anisotropy direction as a free parameter. In particular, one of the latest studies \citep{2024JCAP...11..067A} included higher-degree multipoles in the SSH fit and determined that a substantial amount of cross-talk takes place between the dipole and the quadrupoles, which is caused by the extensive areas of avoidance in the sky distribution of infrared sources (most importantly, the wide exclusion zone along the Galactic plane). Thus, the anisotropy analysis suffers from the same intrinsic problems as our parallax zero-point characterization. 

The updated (and rather uncertain) coordinates of the apparent number density dipole, ranging between $220\degr$ and $260\degr$ in $l$ and $20\degr$ and $40\degr$ in $b$, are curiously close to our estimated direction of parallax offset. \citet{2026MNRAS.546ag248O} tested a few different models representing the anisotropy dipole and the error dispersion and estimated a best-fit direction of $(l,b)=(221\degr,40.8\degr)\pm(11\degr,7.1\degr)$, which is $\simeq 3\sigma$ away from the CMB vector but is nearly perfectly aligned with our negative parallax feature. We surmise that a detailed analysis of the empirical number density with a complete SSH fit will reveal a more complex pattern on the celestial sphere similar to the one we see Fig. \ref{map.fig}. Statistically significant harmonics may reveal a structured character of the cosmic anisotropy. We also considered the number density distribution of our filtered sample of CRF quasars, which does not show any semblance to the dipole of infrared surveys or the parallax offset distribution.

\section{Verification with asteroseismology data}

Although the CRF sample provides the most accurate and detailed map of the sky-correlated parallax bias owing to the large sample size, legitimate concerns arise that the resulting fit can be used for Galactic stars, which may be brighter than the available magnitude range of QSOs, and located in the Galactic exclusion zone, where CRF objects are absent. The collections of parallaxes determined by methods of asteroseismology from precise photometric light curves of red-giant-branch stars \citep{2023A&A...680A.105K} provide an independent test for our correction tool. The authors {\it ibid.} specifically mention that the correction model proposed by \citet{2021A&A...649A...4L} is not suitable for all the fields considered in their study. It is important to see if our method fares better. We used all four tables given in \citet{2023yCat..36770021K} for this verification analysis: 1) Kepler+APOGEE (5307 stars), 2) K2+APOGEE (7997 stars), 3) K2+GALAH (7744 stars), and 4)  TESS+APOGEE (1689 stars). Note that these collections include objects located in separate spots scattered across the celestial sphere, and some of the spots are close to the Galactic equator (Fig. 1 in \citet{2023A&A...680A.105K}).

\begin{figure}
\includegraphics[width=0.52\textwidth]{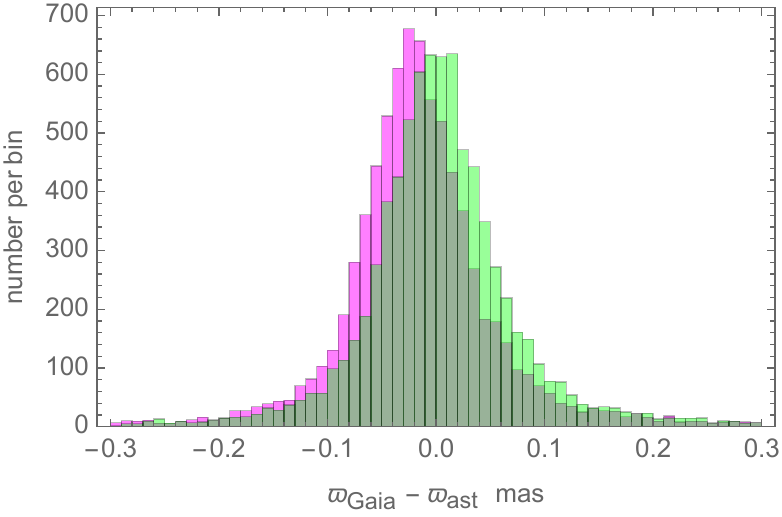}
\caption{Histograms of parallax differences ``Gaia $-$ asteroseismology" for test sample 2 (see text) before (magenta) and after (green) application of bias corrections computed using {\tt varpi.py}.}
\label{astero2.fig}
\end{figure}

We used the {\tt varpi3.py} tool published along with this paper to calculate parallax corrections for each star separately with option {\tt magcor}$=1$, which selects the constant offset coefficient $a_{00}$ according to the input $G$ magnitude. The computed corrections $\varpi(l,b)$ (Eq. \ref{obs.eq}) were subtracted from the parallax differences ``Gaia$-$asteroseismology" given in the input tables. Fig. \ref{astero2.fig} shows the results for one of the fields (\# 2), where the effect of this correction is clearly visible. The original histogram of parallax differences is marked with magenta color, and the sample distribution after the {\tt varpi3.py}-correction with the green color. These distributions are similar in shape, but the obvious negative bias of Gaia parallaxes is practically removed. To quantify the effect of parallax corrections, we determined a uniform fit of the sample distributions for all four fields. We found that they are more faithfully represented by Student T distributions (instead of the expected Gaussians) with three free parameters: mean, scale, and shape. The main parameter of interest is the mean $\mu$, which defines the overall position of the histogram. A negative bias in Gaia DR3 parallaxes corresponds to a negative $\mu$ of the original histogram. Using a suitable Wolfram Mathematica function\footnote{{\tt FindDistributionParameters[data, StudentTDistribution[$\mu$, scale, shape]]}}, we determined that the applied correction changed the means in \uas\ as $-35\rightarrow -18$, $-19\rightarrow +1$, $-20\rightarrow -1$, and $-30\rightarrow -4$ for fields 1--4, respectively. The applied correction also tends to improve the shape parameter $\nu$ of the fitted Student T distributions, which reflects the weight of the non-Gaussian tails: $2.47\rightarrow 2.60$, $1.45\rightarrow 1.96$, $1.70\rightarrow 1.90$, and $1.80\rightarrow 1.80$.

\section{Discussion of possible physical and instrumental effects}

The empirical parallax offset map derived in Sections 3–4, while primarily attributable to instrumental calibration imperfections in the Gaia data reduction, exhibits a suggestive directional alignment with the recently reported quasar number-density dipole. This coincidence raises the question of whether any physical or cosmological effect contributes an intrinsic component to the all-sky negative parallax bias. We first examine whether a cosmological effect, spacetime curvature in a positively curved universe, can contribute (Section~\ref{sec:curv}); we then show that, whatever its origin, the zero-point propagates into the CRF proper-motion field and biases the secular-aberration glide (Section~\ref{sec:pmglide}); and we finally assess a quasi-parallactic component arising from annual orbital aberration (Section~\ref{sec:aber}). Neither physical mechanism reproduces the signal at the required amplitude, leaving instrumental basic-angle variations as the most plausible origin of the parallax bias.

\subsection{Cosmological effects: curvature of spacetime}
\label{sec:curv}
The annual parallax value for distant sources of light is defined by the curvature of the wavefront at the point of measurement. The curvature is inversely proportional to the distance for a spherical wavefront in a flat spacetime. The relation between the distance and the wavefront curvature becomes more interesting in a curved spacetime. In particular, in an isotropic expanding universe with positive curvature $k>0$, the initially spherical wavefront becomes flat (zero parallax) at a specific distance from the source, and then further becomes concave (i.e., curved toward the source away from the observer) at a larger distance \citep[see Appendix A in ][]{2022MNRAS.517.1242M}. The formula for annual parallax in this case can be written as
\eb 
\varpi\equiv 1/D_{\rm par} = \sqrt{k}\,\cot(D_c\,\sqrt{k}),
\ee 
where $D_c$ is the comoving line-of-sight distance, and we used the normalization of the local scale $a(t_0)=1$. The parallax horizon where the parallax distance $D_{\rm par}$ becomes infinity is $D_c=\pi/(2\sqrt{k})$. For more distant sources, a negative annual parallax should be observed.

Available evidence about the basic parameters of the universe indicates that this scenario is not realistic, however. According to the final Planck mission results \citep{2020A&A...641A...6P}, the curvature density parameter value is consistent with zero. Calculating the comoving distance $D_c$ with the evaluated density components $\Omega_m=0.315$ and $\Omega_\Lambda=0.685$, we find that the required curvature density should be $-\Omega_k\equiv k\,D_H^2>0.3$ for the zero-parallax horizon to be within redshift 100. This is totally inconsistent with the current standard model. Furthermore, the Gaia CRF sample includes mostly objects with redshifts less than 3, and we have seen a similar parallactic bias for Galactic stars, which cannot be affected by cosmological effects. Thus, this explanation should be discarded.

\subsection{Propagation of the parallax zero-point into the proper-motion glide}
\label{sec:pmglide}

The negative parallax offset characterized above is usually regarded as a
nuisance confined to the parallax dimension of the astrometric solution. It
is not. Because the parallax $\varpi$ and the two proper-motion components
$\mu_{\alpha*}$, $\mu_\delta$ of each CRF source are estimated jointly from
the same along-scan measurements, they are statistically coupled, and a
systematic offset in the measured parallax propagates into the measured
proper motions through the off-diagonal terms of the covariance matrix. For
extragalactic CRF sources, whose true parallaxes and proper motions are both
zero, this coupling has a direct and quantifiable effect on the vector
spherical harmonic (VSH) determination of the secular-aberration glide.

\begin{figure}[t]
\centering
\includegraphics[width=0.45\linewidth]{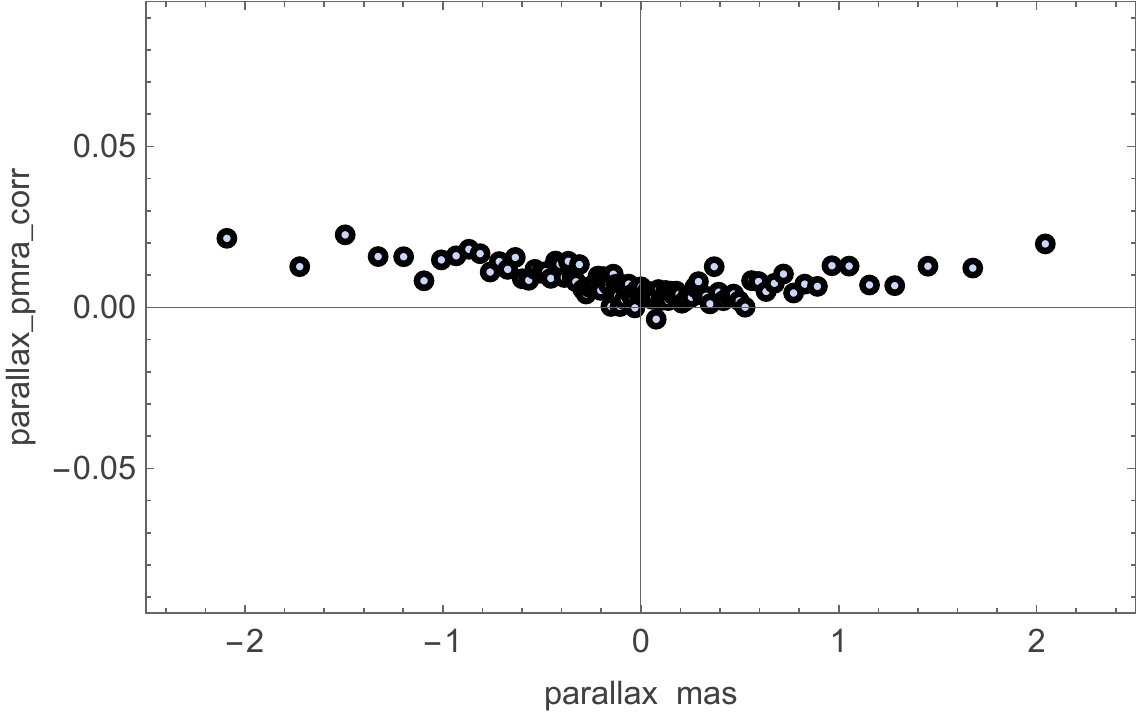}
\includegraphics[width=0.45\linewidth]{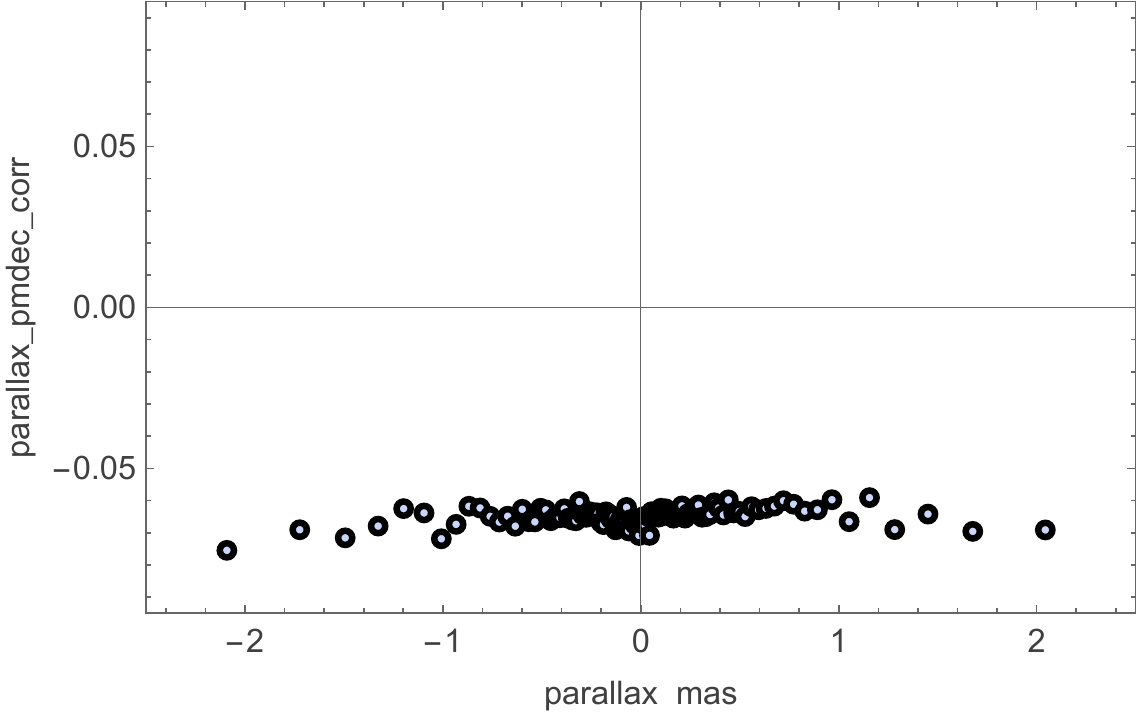}
\caption{Weighted medians of the parallax--proper-motion correlation
coefficients of Gaia CRF sources as a function of measured parallax, in
$101$ equal-population bins. \emph{Left:} the parallax--$\mu_{\alpha*}$
coefficient $\rho_{\varpi\alpha}$ (\texttt{parallax\_pmra\_corr}), which
scatters about zero. \emph{Right:} the parallax--$\mu_{\delta}$ coefficient
$\rho_{\varpi\delta}$ (\texttt{parallax\_pmdec\_corr}), which is flat in
$\varpi$ but displaced to a median of $-0.065$. The decentering of
$\rho_{\varpi\delta}$, combined with the negative parallax zero-point, biases
the declination proper motion through Eq.~(\ref{eq:cond}) and hence the
secular-aberration glide.}
\label{fig:corrpar}
\end{figure}

The relevant covariance descriptors are the parallax--proper-motion
correlation coefficients, denoted \texttt{parallax\_pmra\_corr} and
\texttt{parallax\_pmdec\_corr} in the Gaia archive, which we abbreviate as
$\rho_{\varpi\alpha}$ and $\rho_{\varpi\delta}$. Figure~\ref{fig:corrpar}
shows the weighted medians of these two coefficients for the CRF sample
binned by measured parallax. Two features stand out. First, both
coefficients are essentially independent of $\varpi$ across the entire range
$|\varpi| \lesssim 2$~mas, so that a single, sky-averaged correlation
characterizes the scanning geometry. Second, and decisively, the two
distributions are not centered alike: $\rho_{\varpi\alpha}$ scatters about
zero, whereas $\rho_{\varpi\delta}$ is displaced to a median of $-0.065$.
This decentering of the declination-axis correlation is the channel through
which the parallax zero-point leaks into the proper-motion field.

Consider a CRF source with true parallax and proper motion of zero.
Conditioned on its measured parallax, the expected value of the measured
declination proper motion is the regression of $\mu_\delta$ on $\varpi$
implied by the joint error distribution \citep{Makarov_aposteriori},
\begin{equation}
\langle \mu_\delta \mid \varpi \rangle
   = \rho_{\varpi\delta}\,\frac{\sigma_{\mu_\delta}}{\sigma_\varpi}\,\varpi .
\label{eq:cond}
\end{equation}
The three ingredients of Eq.~(\ref{eq:cond}) are strongly constrained by the
data: the CRF3 sample medians are $\mathrm{med}(\varpi)=-19.4~\mu$as,
$\mathrm{med}(\rho_{\varpi\delta})=-0.065$, and
$\mathrm{med}(\rho_{\varpi\delta}\,\sigma_{\mu_\delta}/\sigma_\varpi)=-0.064$,
the last implying $\sigma_{\mu_\delta}\simeq\sigma_\varpi$ at the median. The
sample median of the right-hand side of Eq.~(\ref{eq:cond}) is driven to zero
when a constant $\Delta\varpi \simeq +25~\mu$as is added to every measured
parallax. This additive shift corresponds to a parallax zero-point of
$\simeq-25~\mu$as, an \emph{internal} determination that invokes no external standard, only the mutual consistency of
$\varpi$, $\mu_\delta$, and $\rho_{\varpi\delta}$ under the zero-parallax
prior, and it agrees, within its scatter, with the direct sample median, with
the SSH constant term $\bar a_{00}$ of Section~\ref{res.sec} and with the
value of \citet{2021A&A...649A...4L}.

The consequence for the reference frame is the point we wish to stress. A
nonzero median $\mu_\delta$ is not a harmless statistic: unlike a rotation or
a random field, the declination-axis proper-motion monopole projects
directly onto the electric first-degree VSH term that carries the
secular-aberration (Galactocentric acceleration) signal. For a glide of
amplitude $g$ toward an apex at declination $\delta_0$, the all-sky mean of
the declination component is
\begin{equation}
\langle \mu_\delta \rangle = \frac{\pi}{4}\,g\,\sin\delta_0 ,
\label{eq:glide}
\end{equation}
(strictly, the coefficient $\pi/4$ holds for uniform sky coverage and is
modified at the few-percent level by the Galactic-plane exclusion of the CRF
sample). The observed CRF median, $\mathrm{med}(\mu_\delta)=-1.7~\mu$as
yr$^{-1}$, is close to the value $-1.9~\mu$as yr$^{-1}$ that
Eq.~(\ref{eq:glide}) returns for the full measured aberration,
$g=5.05~\mu$as yr$^{-1}$ toward the Galactic center
\citep{2021A&A...649A...9G}. The aberration glide and the
parallax-zero-point coupling are therefore \emph{degenerate} at the level of
the median $\mu_\delta$: the same $-1.7~\mu$as yr$^{-1}$ can be read as the
physical dipole or as the propagated parallax bias, and in practice both
contribute. The magnitude of the instrumental term is set by
$|\rho_{\varpi\delta}|\,(\sigma_{\mu_\delta}/\sigma_\varpi)\,|\Delta\varpi|
\sim 1$--$2~\mu$as yr$^{-1}$, i.e., comparable to the aberration itself. A
direct demonstration of this bias is given in the companion analysis
\citep{Makarov_aposteriori},
where removing the conditional coupling
source-by-source shifts the fitted first-degree electric amplitude from
$3.35$ to $5.59~\mu$as yr$^{-1}$.

We therefore conclude that the parallax zero-point enters the VSH glide
determination at the microarcsecond-per-year level, comparable to the
acceleration signal, and must be carried as a systematic uncertainty in any
Galactocentric-acceleration measurement derived from CRF proper motions.
Because the parallax-offset field is itself sky-correlated
(Section~\ref{res.sec}), the same coupling provides a natural, purely
instrumental route to the borderline redshift dependence and the anomalous
low-degree terms reported in recent VSH analyses of the CRF proper-motion
field \citep{2025NatAs...9.1396M}, without recourse to any manipulation of the
delivered proper motions.

\subsection{Annual aberration}
\label{sec:aber}
Relativistic aberration of light is removed from Gaia observational data at the early processing steps. It seems to be an unlikely source of parallax error, because the position perturbation dipole caused by aberration is aligned with the instantaneous velocity vector of the observer at the time of observation. Since the observer's velocity vector is nearly orthogonal to the sun direction, the aberrational shift of the observed position is nearly orthogonal to the nominal parallactic displacement (which is directed almost exactly toward the sun). In other words, the parallactic and aberrational effects are almost orthogonal on the sphere. The key word is ``almost", and it is easy to demonstrate that a small systematic error in the observer's ephemerides can generate a dipole distribution of parallax error.

The amplitude of annual aberration is approximately $20.5\arcsec$. It is caused by the orbital velocity of Earth around the barycenter of the Solar system (which is slightly shifted from the solar center) of roughly 30 \kms. The smooth component of the observed parallax bias has an amplitude of approximately 20 \uas, which is one part in a million relatively to the nominal aberration amplitude. However, since the main aberration component is orthogonal to the parallactic displacement, we should only consider the radial component of observer's barycentric velocity, which is a small fraction of the tangential velocity. This component is, specifically,
\eb 
v_r=\frac{nae\,\sin{E}}{1-e\,\cos{E}},
\ee 
where $n$ is the mean motion, $a$ is the semimajor axis, $e$ is the orbital eccentricity, and $E$ is the eccentric anomaly. The extrema of $v_r$ are reached at $E=\arccos(e)$ (maximum) and $E=2\pi-\arccos(e)$ (minimum). The maximal outward velocity is
\eb 
v_r^{\rm max}=\frac{nae}{\sqrt{1-e^2}}.
\ee 
For a small eccentricity, the ratio of the maximal radial velocity to the total mean velocity approximately equals $e$. The eccentricity of Earth is approximately 0.0167. The mean anomaly of the point of maximum radial velocity equals $\arccos(e)-e\sqrt{1-e^2}$, which simplifies if expanded in powers of eccentricity to $\pi/2-2e$. More importantly for the present study, using the well-known exact relation between true anomaly $\nu$ and eccentric anomaly $E$, we establish that the
true anomaly of the point where $v_r=v_r^{\rm max}$ is exactly $\pi/2$ (semilatus rectum). The current ecliptic longitude of Earth perihelion is $\simeq 103\degr$, thus, the longitude of maximal radial velocity is $\simeq 193\degr$. If the true eccentricity is larger by 0.00006 than the nominal assumed value, the sources on the meridian at ecliptic longitudes $103\degr$ and $283\degr$ would be seen to have a shift toward the semilatus rectum direction by 20 \uas\ resembling a {\it negative} parallactic effect. However, since the radial velocity variation is anti-symmetric around the line of apsides, a positive pseudo-parallactic effect emerges when the same source is observed with the opposite sun direction. The maximal aberration-related displacement cannot be achieved for Gaia's modus operandi, because the angle between the space craft's rotation axis and the sun is always kept at $45\degr$.\footnote{{\url https://gea.esac.esa.int/archive/documentation/GDR3/Introduction/chap\_cu0int/cu0int\_sec\_mission/cu0int\_ssec\_scanning\_law\_concepts.html}} This effectively reduces the quasi-parallactic shift by a factor $1/\sqrt{2}$. 

However, these considerations are of rather scholastic nature for the following reasons. Gaia is not attached to Earth but moves around the L2 point in a quasi-periodic Lissajous orbit with a period close to 180 d. Earth eccentricity, which is an osculating and secularly varying element, is not used in the determination of Gaia's barycentric velocity. Instead, the barycentric location and velocity are computed using a combination of the numerical Earth ephemeris, Earth orientation parameters, and active radio tracking measurements with dedicated antennae on the ground. All these components are believed to have superior accuracy meeting the mission requirement of 0.1 mm s$^{-1}$. The spacecraft does constantly move outward under the pressure of solar light, but this effect should be accurately measured and calibrated out. Technically, the weakest link in the control system is the ionospheric plasma density correction of the Doppler-shifted tracking signal, which has to rely on a single frequency X-band. Still, it appears unlikely that the ephemeris control system would miss a constant radial drift of $\sim 30$ mm s$^{-1}$ that could cause the observed all-sky bias of parallax.

\section{Conclusion}

We have introduced a practical method to characterize and correct the sky-correlated, magnitude-dependent parallax bias of Gaia DR3, based on a scalar spherical harmonic decomposition to degree 8 of the measured parallaxes of $\sim\!10^6$ CRF quasars, and released it as the Python tool {\tt varpi3.py}. Of the 81 harmonic terms, only the constant $Y_{00}$ term depends significantly on $G$ magnitude, ranging from $\simeq-18$ \uas\ at the bright end to near zero at $G\simeq20.7$, while the remaining position-dependent terms are magnitude-independent within the errors. The reconstructed offset map is dominated by a dipole-like structure lying close to the ecliptic plane, whose most negative pole, $(l,b)\simeq(220\degr,+43\degr)$, is intriguingly aligned with the reported quasar number-density dipole. Verification against asteroseismic parallaxes of red-giant stars in four independent fields confirms that the correction removes most of the negative bias even for objects brighter than the CRF magnitude range and located in the Galactic exclusion zone. Of the physical mechanisms considered, neither positive spatial curvature nor annual orbital aberration can reproduce the signal at the required amplitude, leaving instrumental basic-angle variations as the most plausible origin.

The imminent Gaia data release 4 (DR4) is expected to provide a larger CRF sample with much improved precision of proper motions and parallaxes. Since the drivers of the sky-correlated and generally negative parallax offset have not been identified, a similar problem may be present in DR4. Our method can be easily adapted for the new data set. Possible upgrades include higher-degree SSH fits capitalizing on the larger sample, more detailed characterization of the position-dependent component, and specific propagation routine for parallax-correlated proper motion and position corrections.

One of the outstanding issues of the present technique {\tt varpi3.py} is its limited range on the bright end, which is defined by the bright cutoff of QSO objects. Assuming that Gaia DR4 parallax offset carries the same feature as in DR3, the magnitude-dependent constant term $Y_{00}$ can be differentially determined at brighter magnitudes using sufficiently compact and distant aggregates of stars, such as globular clusters. We also note that the CRF magnitudes themselves are apparently correlated in DR3 with the astrometric errors ellipse properties \citep{2026RNAAS..10..201M}---hence, the empirical higher-order harmonics of the SSH fit may be caused by the non-uniform scanning pattern and corresponding anisotropies of the astrometric uncertainty. This investigation may be a step toward a unified model of Gaia systematics.

Finally, we note a consequence of the parallax bias that reaches beyond
distance determinations. Because parallax and proper motion are estimated
jointly, the negative zero-point propagates into the proper motions of the
CRF sources through the actual decentered parallax--proper-motion correlations; as
shown in Section~\ref{sec:pmglide}, this produces a median declination
proper motion of $-1.7~\mu$as~yr$^{-1}$ that projects directly onto the
first-degree electric harmonic carrying the secular aberration. The
parallax zero-point therefore enters the astrometric determination of the
Galactocentric acceleration \citep{2021A&A...649A...9G} at the
microarcsecond-per-year level, comparable to the signal itself, while the
mutual consistency of $\varpi$, $\mu_\delta$, and their correlation
furnishes an independent estimate of the offset ($\simeq -25~\mu$as) that
corroborates the SSH value reported here. A dependable parallax correction
is thus a prerequisite not only for the distance scale but also for the
reference-frame and cosmological uses of the Gaia CRF proper-motion field.

\section{Acknowledgments}
This work supports USNO's ongoing research into the celestial reference frame and geodesy. 
The authors thank Oleg Titov for inspiring discussions of possible physical effects resulting in negative parallax measurements, and to Claus Fabricius for sharing his insight in Gaia mission operations.

\vspace{5mm}
\facilities{Gaia}


\software{
    \texttt{Mathematica} \citep{Mathematica}, Python
    }

\bibliography{main}{}
\bibliographystyle{aasjournal}

\section*{Appendix}
The Python-based {\tt varpi3.py} tool published along with this paper can be used to compute the empirically estimated zero-point parallax offset in Gaia DR3 for a source at any position on the sky. The user-provided input is a list of at least one position in the ICRS (equatorial) system defined by its RA and Decl. coordinates in decimal degrees and, optionally, the source's Gaia $G$ magnitude. The algorithm performs the following operations. 
\begin{enumerate}
    \item 
The input position is transformed to the standard galactic coordinate system. 
\item 
A table of 81 SSH coefficients pre-computed by us is ingested, which defines the position-dependent distribution of parallax offset.
\item 
If a magnitude is given in the user's input as a third column, a predefined sequence of $a_{00}(G)$ nodal values is used to interpolate or extrapolate the closest value of the first SSH coefficient, effectively performing a zeroth-order interpolation on the grid of 10 fixed nodes. The selected coefficient can thus have one of the 10 fixed values, and all sources brighter than $G=18.384$ or fainter than $G=20.681$ mag will be assigned the same $a_{00}$ coefficients corresponding to the end-points of the sequence.
\item 
An SSH series to degree $N=8$ is computed for each input data point. These values in \uas\ should be subtracted from the Gaia parallaxes given in the catalog.
\item 
If the input table of positions represents a sufficiently dense grid covering the entire sky, an optional graphical function is provided to depict the SSH-fitted parallax distribution in a galactic Mollweide projection of the celestial sphere.

\end{enumerate}
\end{document}